\documentclass[numsec,webpdf,modern,medium,namedate]{article}

\onecolumn

\usepackage{amsthm}
\theoremstyle{thmstyleone}%
\newtheorem{theorem}{Theorem}
\theoremstyle{thmstyletwo}%
\theoremstyle{thmstylethree}%

\usepackage[doublespacing]{setspace}
\usepackage[fontsize=12pt]{fontsize}

\PassOptionsToPackage{shortlabels}{enumitem}

\usepackage{rasmus2,natbib,color}
\usepackage{amssymb}
\usepackage{xcolor}
\usepackage{hyperref}
\usepackage{graphicx}
\usepackage{xr}
\usepackage{caption}
\hypersetup{
    colorlinks,
    linkcolor={blue!50!black},
    citecolor={blue!50!black},
    urlcolor={blue!80!black}
}
\definecolor{green2}{HTML}{008000}

\newcommand{\M}{\mathcal M}
\newcommand{\CC}{\mathcal C}
\newcommand{\RR}{\mathbb R}

\newcommand{\norm}[1]{\left\lVert#1\right\rVert}

\DeclareCaptionLabelFormat{myformat}{\fontsize{10}{12}\selectfont#1~#2}
\usepackage{geometry}
\usepackage{authblk}
\title{Semi-parametric Markov models for multi-type point patterns}
\author[1]{Ib Thorsgaard Jensen}
\author[2]{Jean-Fran\c{c}ois Coeurjolly}
\author[1]{Rasmus Waagepetersen}
\affil[1]{Department of Mathematical Sciences, Aalborg University, Thomas Mannsvej, 9220 Aalborg {\O}st, Denmark}
\affil[2]{Laboratory Jean Kuntzmann, Universit{\'e} Grenobles Alpes, 150 Place du Torrent, 38400 Saint-Martin d'Hères, France}
\date{}                     
\begin{document}
%
%
%
%
%



\maketitle
\abstract{Multi-type Markov point processes offer a flexible framework for modelling
  complex multi-type point patterns where it is pertinent to  capture
  both interactions between points as well as large scale trends
  depending on observed covariates. However, estimation of interaction
  and covariate effects may be seriously biased in the presence of
  unobserved spatial confounders. In this paper we introduce a
  new class of semi-parametric Markov point processes that adjusts for
  spatial confounding through a non-parametric factor that
  accommodates effects of latent spatial variables common to all types
  of points. We introduce a
  conditional pseudo likelihood for parameter estimation and
  show that the resulting estimator has desirable
  asymptotic properties. Our methodology not least has great potential in studies of
  industry agglomeration and we apply it to study spatial patterns of
  locations of two types of banks in France.}


\section{Introduction}

Spatial multi-type (or multivariate) point pattern data where points represent objects or events of
different types are abundant. Some examples of data sets are locations of
rain forest trees of different species, crime scenes of different
types of crimes, and locations of different types of industries or banks. The main interests for
such data is to study possible dependence on spatial
covariates and within- or between-type interactions between points. The statistical toolbox
for this can roughly be divided into non-parametric
cross summary
statistics \citep[e.g.][]{moeller:waagepetersen:03,baddeley:jammalamadaka:nair:14,cronie:lieshout:16}
and parametric methods.

Parametric
modelling may comprise only the first and second order properties of a
multi-type point process, more specifically the intensity function
and the cross pair correlation functions
\citep{moeller:waagepetersen:03}. It could also involve specification of the full multi-type point process
distribution in terms of  multi-type cluster processes
\citep{jalilian:etal:15}, multi-type log Gaussian Cox processes
\citep{brix2001space,waagepetersen:etal:15}, hierarchical Markov point
processes \citep{hogmander:sarkka:99,grabarnik:sarkka:09}, or multi-type Markov point processes
\citep{picard2009multi,goulard1996parameter,coeurjolly:etal:12,rajala:murrell:olhede:17}. 
Multi-type Markov point processes are typically interpreted in terms
of the so-called Papangelou conditional intensity which essentially
models the probability of occurrence of an event or object of a
specific type and at a specific location conditional on an existing
multi-type point configuration. The conditional intensity
simultaneously comprises the effects of large scale trends
(inhomogeneity) and interactions.

In this paper we introduce semi-parametric inference for multi-type
Markov point processes. The basic structure of our proposed model is similar to the
one considered in \cite{rajala:murrell:olhede:17} but crucially, we
enrich the conditional intensity with a non-parametric factor. This factor is common to all types of points and
allows the modelling of large scale variation due to unobserved
possibly confounding spatial variables. In the context of criminology or
spatial econometrics, confounding variation could e.g.\ be caused by a
varying population density which may be hard to quantify at a fine
spatial resolution or by a complex urban environment. The problem of spatial confounding has received considerable interest in the context of random field data \cite[e.g.][]{dupont2022spatial+} but much less so for spatial point pattern data which further supports the importance of our contribution. Due to the presence of the non-parametric factor,
  standard pseudo likelihood estimation \cite[e.g.][]{goulard1996parameter} is not applicable. We therefore  propose a conditional pseudo likelihood function that does not  depend on the non-parametric factor.  
  
Our proposed model and estimation method is related to  \cite{hessellund:etal:22a} who introduced
semi-parametric models for the intensity functions of a multi-type
point process and a composite multinomial logistic regression object function for parameter
estimation.  However, compared with
\cite{hessellund:etal:22a} a distinct advantage of the new method is
that first and second order properties can be inferred
simultaneously. In contrast, the method in \cite{hessellund:etal:22a}
for estimating the intensity function (i.e.\ first order properties)
needs to be supplemented by a second step for inferring second-order
properties \citep{hessellund:etal:22b}. 
The method in \cite{hessellund:etal:22b} further does not work for  multi-type point processes with repulsion between points of the same type. This is e.g.\ a serious restriction in relation to our data example.

We derive asymptotic properties of the conditional pseudo likelihood parameter estimator in the setting of increasing spatial
domain. Since a parametric model for a multi-type Markov point process
is a special case of the proposed semi-parametric model (with the
non-parametric factor equal to one), we fill a gap in the existing
literature on multi-type inhomogeneous Markov point processes, as the
asymptotic results in  \citet{coeurjolly:etal:12}
and~\citet{ba:coeurjolly:23}  are limited to respectively the
  stationary or the univariate case. Finite sample properties of parameter estimates and
confidence intervals are investigated in a simulation study.

\begin{figure}[t]
\caption{{\fontsize{10pt}{12pt}\selectfont Locations of two types of French banks.}}
  \centering
 \label{fig:Banks}
 \includegraphics[width=0.8\textwidth]{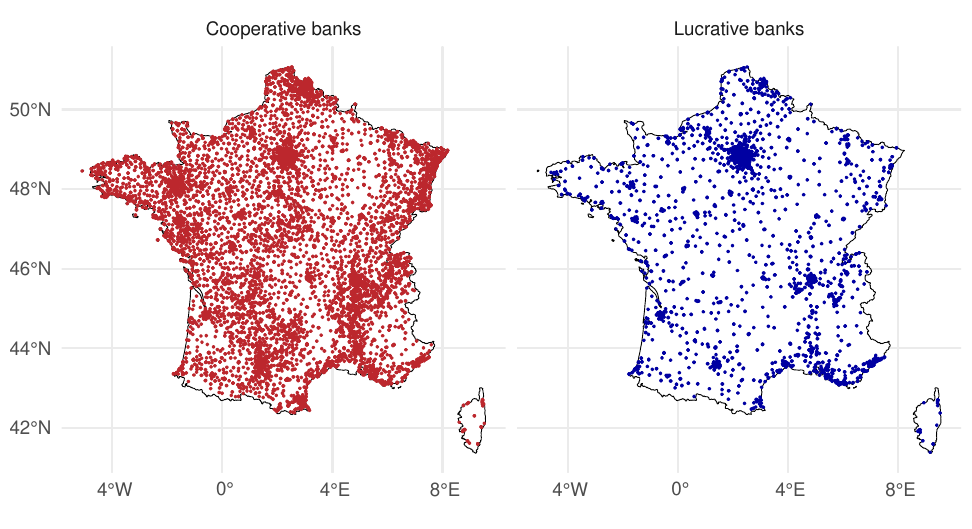}
 \vspace*{-1cm}
\end{figure}

Our approach is strongly motivated by studies of industry
  clustering in spatial econometrics. Without explicitly
mentioning point processes, the highly influential paper
\cite{duranton:overman:05} acknowledged the need
to take into account common large scale trends when studying
clustering of different types of industries in terms of non-parametric
methods. This is in line with \cite{hoover:37} who defined industry
co-localization as the agglomeration of an industry after controlling for
general manufacturing. Later, \cite{sweeney:gomez-antonio:16} and \cite{gomez-antonio:sweeney:21} applied
univariate Markov point process models to study within-type industry interactions as well as dependence on various
economically relevant covariates. However, these papers used purely parametric methods
and did not consider multi-type aspects of industry
clustering. Our paper therefore also fills a gap in the spatial
econometrics literature. We demonstrate the usefulness of our approach
by a study of within and between interaction for a bivariate
point pattern of so-called lucrative and cooperative banks in France,
see Figure~\ref{fig:Banks} and Section~\ref{sec:banks}.

\section{Background on point processes and multi-type Markov point
  processes}

A multi-type point process
$Y$ on $\R^d$ is a random locally finite set of marked points $(u,m)\in \R^d
\times \mathcal M$ where the mark space $\mathcal M$ is finite. Specifically we let $\mathcal M=\{1,\ldots,p\}$ for some $p \ge 1$. Letting $Y_B= Y \cap (B
\times \mathcal M)$ denote the marked points $(u,i)$ in $Y$ whose
location part $u$ falls in $B \subseteq \R^d$, local finiteness of $Y$
means that $Y_B$ is finite almost surely for any bounded $B\subset
\R^d$. We define in the same way $y_B$ for $y \subseteq \R^d \times
\mathcal{M}$ and denote by $\mathcal C$ the set of all locally finite
configurations and by $\mathcal C_{B}$ the restriction of $\mathcal C$
to configurations on $B \subseteq \R^d$, i.e.\ to $y \in \mathcal C$
with $y \subset B \times \mathcal M$. For $B \subseteq \R^d$, $|B|$ denotes the volume of $B$.

A multi-type point process $Y$ can be identified with a multivariate point
process  $(X_1,\dots,X_p)^\top$ where $X_i$ consists of those points $u$ for which $(u,i) \in Y$. Moreover, $X=\cup_{i=1}^p X_i$ is the point
process of all points in $Y$ with their marks removed. Similarly we use the generic notation $x_i=\{u \in \R^d|(u,i) \in y\}$ for the set of type $i$ points in a multi-type configuration $y$. For any
  $B \subseteq \R$ we let $N_{X_i}(B)$  and $N_X(B)$ denote the number of points in
$X_i$ respectively $X$ that fall in $B$ (i.e.\
the numbers of points of type $i$ respectively of any type in $Y_B$).

In the following sections we consider finite or infinite multi-type point processes specified in terms of a density function or the so-called Papangelou conditional intensity. To avoid excessive use of braces, we abuse notation and write $y \cup (u,i)$ for $y \cup \{(u,i)\}$ and $y\setminus (u,i)$ for $y \setminus  \{(u,i)\}$.
\vspace*{-0.8cm}
\subsection{Finite multi-type point processes}

 For a bounded $B \subset \R^d$, we consider finite multi-type  point processes on $S=B \times
\mathcal M$ 
specified by a density $f: \CC_B \rightarrow [0,\infty [$ with respect
to the unit rate multi-type Poisson point process on $S$.
The density is of the form
\begin{equation}\label{densityGibbs}
 f(y) = \frac{1}{c} g(y), \quad y \in \CC_{B},
\end{equation}
where $g$ is an unnormalized density defined as a function $g: \CC \rightarrow
[0,\infty]$ with the property that $g(y)$ is finite when $y$ is
finite. In Section~\ref{sec:semiparametricmodel} we specify $g$ in
more detail as a product of certain non-negative interaction functions. The
normalizing constant $c$ is usually
intractable. The reason for using the domain $\CC$ for $g$ will become
clear in Section~\ref{sec:infinitegibbs} where we discuss infinite
multi-type  point processes on $\R^d$ and want to avoid
  reference to a specific $B$.

We exclusively consider the situation where $g$ (or equivalently $f$) is hereditary, that is, $g(y)>0  \Rightarrow
g(y^\prime) >0$ for finite $y \in \mathcal{C}$ and $y^\prime \subset y$. Hence any subset $y'$ of an
allowed finite configuration $y$ is allowed too. Then for any finite $y \in \mathcal C$ and $(u,i)\in S$  with $g\{y \cup (u,i)\}>0$, the Papangelou conditional intensity is well-defined as
\begin{equation}\label{eq:PapangelouMarked} \ld\{(u,i),y \} = \frac{g\{y\cup (u,i)\} }{g\{y \setminus
    (u,i)\}} 
\end{equation}
and heuristically, the quantity $\ld\{(u,i),y \}
\dd u$  may be interpreted as the probability of
observing a marked point with mark $i$ in an infinitesimal
neighbourhood of $u$ given that $Y$ outside this neighbourhood
coincides with $y$. The definition of $\ld$ implies that $\ld\{(u,i),y
\}=\ld\{(u,i),y \setminus (u,i)\}$ when $(u,i) \in y$. For the set of
$y$ and $(u,i)$ with $g\{y \cup(u,i)\}=0$ we arbitrarily define
$\ld\{(u,i),y\}=0$.

The Papangelou conditional
intensity is said to be locally stable (LS) if there exists $\bar \lambda<\infty$ such that for any $(u,i)\in S$ and finite $y \in \mathcal C$,
\begin{equation}
\label{eq:LS}
 \ld \{ (u,i) , y\} \leq \bar \lambda \tag{LS}.
\end{equation}
Local stability \eqref{eq:LS} ensures that $g$ has finite integral $c$
over $\CC_B$ so that the density $f$ is well-defined \cite[e.g.][]{moeller:waagepetersen:03}.

\subsection{Multi-type point processes on $\R^d$}\label{sec:infinitegibbs}

When $S=\R^d\times \mathcal M$, the density~\eqref{densityGibbs} does
not make sense in general since $g(y)$ could be infinite for $y \in C
\subset \CC $ where $C$ has positive probability under the unit rate
multi-type Poisson process. Instead we extend the definition of the Papangelou conditional intensity
\eqref{eq:PapangelouMarked} to $y \in \CC$. 

To ensure the validity of the definition \eqref{eq:PapangelouMarked}
for infinite configurations $y$, we introduce a finite range property (also known as a Markov property),
$\lim_{r \rightarrow \infty} g\{y_{B(u,r)} \cup (u,i)\}/g\{y_{B(u,r)}\} = g\{ y_{B(u,R)} \cup (u,i) \}/g\{y_{B(u,R)}\}$
for some $R < \infty$ where $B(u,R)$ denotes the
Euclidean ball with center $u$ and radius $R$. The Papangelou
conditional intensity is then well-defined in the infinite case as
\begin{equation}
\label{eq:FR}
 \ld\{(u,i),y\}=\ld\left\{(u,i) , y \cap B(u,R)\times \mathcal M \right\} \tag{FR}
\end{equation}
where the right hand side is obtained from \eqref{eq:PapangelouMarked}.
In other words, the probability of observing a marked point $(u,i)$
given a surrounding configuration $y$ only depends on the
$R$-neighbours of $u$ in $y$. Multi-type point processes satisfying this condition are examples of multi-type Markov point processes \cite[][]{moeller:waagepetersen:03}.

 We then say that $Y$ is a multi-type point process on $\R^d$ with
conditional intensity of the form \eqref{eq:FR} for $y
\in \CC$ provided
\begin{equation}
\label{eq:GNZ}
\EE \sum_{(u,i) \in Y} h\{ (u,i),Y \setminus (u,i) \}  =
 \EE \; \sum_{i=1}^p\int    h\{(u,i) , Y  \} \ld \{ (u,i) , Y\}  \mathrm{d}u
\end{equation}
for any non-negative function $h : (\R^d \times \mathcal M)\times
\mathcal C \to [0,\infty[$.  The equation \eqref{eq:GNZ} is the
so-called GNZ
formula~\citep{xanh:zessin:79,georgii:79,georgii:11} for multi-type
 point processes. It is of course also valid for a finite
multi-type point process as considered in the previous section. 

Given the definitions of finite range and local stability we can state
the following existence result.

\begin{theorem}{[\citet[Theorem~5.6]{jansen2018gibbsian}]}
\label{thm:existence}
Suppose the Papangelou
conditional intensity $\ld : (\mathbb{R}^{d}\times \mathcal
M) \times \mathcal C \to  [0,\infty[$ specified by
\eqref{eq:PapangelouMarked} satisfies \eqref{eq:FR} and \eqref{eq:LS}. Then there exists at least one infinite volume multi-type point process $Y$  with Papangelou conditional intensity $\ld$ satisfying the GNZ equation~\eqref{eq:GNZ}.
\end{theorem}

There may exist distinct infinite volume point processes sharing the same Papangelou conditional intensity but  their
conditional distributions on bounded $B$ agree. For an infinite volume point process $Y$ the 
conditional distribution of $Y_B$ given $Y_{B^c}=z$, $z \in \CC_{B^c}$,
is specified by a  density of the form
\begin{equation}\label{eq:conddens} f_B(y | z) \propto g(y \cup z), \quad y \in \CC_B \end{equation}
which is closely related to the finite case density
\eqref{densityGibbs}. 

In addition to the GNZ formula we also use the iterated GNZ formula 
\begin{align}\label{eq:iteratedGNZ}
 & \EE \sum_{(u,i),(v,j) \in Y}^{\neq} h[ (u,i),(v,j),Y\setminus \{(u,i),(v,j)\} ] \nonumber \\ = &\EE \sum_{i,j=1}^p\int h\{(u,i) ,(v,j), Y  \} \ld \{ (u,i) ,(v,j), Y\}  \mathrm{d}u \dd v
\end{align}
for any non-negative function $h:\{(\R^d \times \mathcal M) \times
\mathcal C\}^2 \to [0,\infty[$ and where $\ld \{ (u,i) ,(v,j), Y\}= \ld \{ (u,i), Y\}\ld \{ (v,j), Y \cup (u,i) \}$.

\section{Semi-parametric modelling and methodology}

\subsection{Semi-parametric Markov point processes}\label{sec:semiparametricmodel}

We introduce a specific class of multi-type Markov point processes
with points in a bounded domain $W \subset \R^d$ by specifying a particular model for the unnormalised density $g$ in~\eqref{densityGibbs}. We consider densities $f$ of the form
\begin{equation}\label{eq:densityMarked}
f(y) \propto g(y)= \prod_{(u,i) \in y} \phi_i(u)\prod_{i,j=1}^p
  \phi_{ij}(x_i,x_{j}),
\end{equation}
where $y \in \mathcal C_W$ is any  configuration of marked points in
$W \times \mathcal M$. The non-negative functions $\phi_i$ are used to model first order effects, i.e.\ spatial trends in the occurrence of type $i$ points, and are assumed to be of the form
\begin{equation}\label{eq:intensity}
   \phi_i(u)=\phi_0(u) \exp\{\gm_{i0}^\top z_i(u)\}
\end{equation}
where $z_i(u)$ and $\gamma_{i0}$ respectively denote a vector of covariates at
location $u$ and a regression parameter vector. The function $\phi_0$
is common to all $X_i$ and is unrestricted except for being
non-negative. This function constitutes a key component of our model
  since it can capture large scale variation common to all types of
  points yet not explained by the observed covariates.

The non-negative functions $\phi_{ij}$ model higher-order interaction terms. We
assume that these interactions take the following log-linear  form
\begin{equation}\label{eq:interaction}
\phi_{ij}(x_i,x_j) = \exp\{ \gm_{ij}^\top v_{ij}(x_i,x_j) \}
\end{equation}
where the $v_{ij}$
are (possibly) vector-valued functions depending on pairs of point patterns and
the $\gm_{ij}$ are interaction parameter vectors. Our examples of $v_{ij}$ below depend on an interaction range $R_{ij}$ which may be considered known or a further target for estimation. 
\begin{itemize}
\item Multi-type Strauss hard-core interaction: Define for  interaction ranges
$0\le R_{ij}$,
\[ s(u,x_j
    ,R_{ij})= \left(\frac{1}{2}\right)^{1(i=j)} \sum_{v \in x_j\setminus u}
    1( \|u -v
    \|\le R_{ij}) \]
to be the number of $R_{ij}$ close neighbours of $u$ in $x_j \setminus u$.
Introducing further hard-core distances $0 \le r_{ij} \le R_{ij}$,
\[
    v_{ij}(x_i,x_j)= \begin{cases}
    \sum_{u \in x_i} s(u,x_j ,R_{ij})  & \text{ if }\|u-v\| \ge r_{ij}\text{ for all }  u \in x_i, v \in x_j \setminus u \\
  -\infty & \text{otherwise}.
\end{cases}
\]
With $r_{ij}=0$ for all $i,j$, a multi-type Strauss interaction is obtained.
\item Multi-type Geyer saturation interaction: \[v_{ij}(x_i,x_j)=\sum_{u
      \in x_i} \min \{
    s(u,x_j,R_{ij}),c_{ij}\}\]
where $s(u,x_j,R_{ij})$ is defined as in the previous example and
$0<c_{ij} <\infty$ is the saturation parameter.
\end{itemize}

In the homogeneous case, i.e.\ when the $\phi_i$ are constant, it is known \citep[see
e.g.][]{baddeley:rubak:turner:15,ba:coeurjolly:23}, that the multi-type
Strauss hard-core point process (with $r_{ij}>0$ or $\gamma_{ij} <0$ in
case $r_{ij}=0$) and the multi-type Geyer satisfy~\eqref{eq:LS} and~\eqref{eq:FR}
with respectively $R=R^\vee,2R^\vee,$ where
$R^\vee=\max_{i,j}\{ R_{ij}\}$. If one adds simple
  conditions like $\sup_i \|z_i\|_{\infty}< \infty$ and $\|\phi_0\|_\infty < \infty$, the inhomogeneous versions of these models still satisfy~\eqref{eq:FR}-\eqref{eq:LS} and so exist in the infinite volume according to Theorem~\ref{thm:existence}. In addition to multi-type Strauss and multi-type Geyer, other options also exist such as the multi-type area interaction model \citep{picard2009multi,nightingale2019area}.

Equations~\eqref{eq:intensity}-\eqref{eq:interaction} allow us to rewrite the density as
\[
  f(y) \propto \left\{\prod_{(u,i)\in y} \phi_0(u) \right\}\;  \exp \left\{ \gm^\top v(y)   \right\}
\]
where $\gm$ and $v(y) $ are the stacked vectors given by
$\gm = ( \gm_{1}^\top,\ldots,\gm_p^\top)^\top$ and\\ $v(y) = \{v_1(y)^\top,\ldots,v_p(y)^\top\}^\top$
with
$\gm_{i}= (\gm_{i0}^\top,\gm_{i1}^\top,\ldots,\gm_{ip}^\top)^\top$ and\\
$$v_i(y) = \left \{ \sum_{u\in x_i}
         z_i(u)^\top,v_{i1}(x_i,x_1)^\top,\ldots,v_{ip}(x_i,x_p)^\top
         \right \}^\top.$$ Then the Papangelou conditional intensity~\eqref{eq:PapangelouMarked} of $Y$ becomes
\begin{equation} \label{eq:PapangelouMarked2}
\ld\{(u,i),y \}  = \phi_0(u) \; \exp \left[ \gm^\top v\{(u,i),y \}  \right]
\end{equation}
for $(u,i)\in \R^d \times \mathcal M$ and $y\in \mathcal C$, where
we define
  \begin{equation}\label{eq:localcontribution} v\{(u,i),y \} = v\{y\cup(u,i)\}-v\{y\setminus(u,i)\} \end{equation}
  which measures the local contribution of a marked point $(u,i)$ to a configuration $y\setminus (u,i)$ or $y$  depending on whether $(u,i) \in y$ or not. We define for $(u,i) \in \R^d \times \mathcal{M}$ and $y\in \mathcal C$ the following probability that does not depend on $\phi_0$,
\begin{equation} \label{eq:p}
p\{ (u,i),y\} = \frac{\ld\{ (u,i),y\}}{ \Ld (u,y) }  =\frac{\exp \left[ \gm^\top v\{(u,i),y\} \right]}{\sum_{j=1}^p\exp \left[ \gm^\top v\{(u,j),y\} \right]},
\end{equation}
where $\Ld (u,y)  = \sum_{j=1}^p \ld \{ (u,j),y\}$ and where we take the convention $0/0=0$ in~\eqref{eq:p}.

By the infinitesimal interpretation of $\ld \{(u,i),y)\}$,
$\Ld(u,y)\dd u$ can be interpreted as the conditional probability of
observing a marked point of {\em any} type in the vicinity of
$u$ given the rest of the point configuration. It follows that $p\{ (u,i),y\}$ can be interpreted as the
probability that a point located at $u$ is of type $i$
given that a point from $Y$ occurs at $u$ and given that $Y$ coincides
with $y$ in the surrounding of $u$.

\subsection{Conditional pseudo likelihood}

Assume we have a multi-type Markov point process model
satisfying~\eqref{eq:LS}-\eqref{eq:FR} observed on $W$. Since the
density \eqref{eq:densityMarked} is only specified up to an unknown
normalizing constant, maximum likelihood estimation is not
straightforward. The so-called pseudo likelihood \cite[e.g.][]{goulard1996parameter} formulated in terms of the Papangelou
conditional intensity is therefore often employed instead. The
conditional intensity does not depend
on the normalizing constant but in our case the presence of the
unspecified non-parametric component rules out the usual
pseudo likelihood approach. Instead we propose a conditional pseudo
likelihood approach as detailed in the following.

Let $D=W\ominus R$, which is the domain $W$ eroded by $R$. It follows
from \eqref{eq:FR} that for any $(u,i)\in D\times \mathcal M$ and $y\in \mathcal C$, $\lambda\{(u,i) , y\}$ and therefore $p\{(u,i) , y\}$ can always be evaluated using $y_W$. To estimate the parameter vector $\gm$, we propose to use the conditional pseudo likelihood
$L(\gm) = \prod_{(u,i)\in Y_D} p\{(u,i) , Y\setminus (u,i)\}.$
The conditional pseudo log-likelihood $\ell(\gm) = \log L(\gm) $ is
\begin{equation} \label{eq:loglik}
\ell(\gm) = \sum_{(u,i)\in Y_D} \Big[
 \; \log \ld\{(u,i), Y \setminus (u,i) \}
- \log  \Ld\{u, Y \setminus (u,i)\}   \;
\Big].
\end{equation}
With the logistic form~\eqref{eq:p} for the probabilities $p$,
$\ell(\gm)$ is equivalent to the log
likelihood of a multinomial logistic
regression. However, in a standard
multinomial logistic regression, the same set of covariates is usually
shared among the $p$ types, which requires $z_i=z_j$ and
$v_{il}(\cdot,\cdot)=v_{jl}(\cdot,\cdot)$ for $i\neq j$ and each $l=1,\ldots,p$.
In this case parameter estimates can be
obtained easily using standard software for multinomial logistic regression,
e.g.\ the \texttt{VGAM} \texttt{R} package.  However, if different
covariates are used for different types or $v_{il}(\cdot,\cdot)\neq
v_{jl}(\cdot,\cdot)$ for an $l$ and some $i \neq j$, tailor made
code is necessary. Also certain identifiability issues may necessitate reparametrizations, see Sections~\ref{sec:ident} and \ref{sec:reparm}.

\subsection{Identifiability}
\label{sec:ident}
For the interaction components of $v\{(u,i),y\}$ given by
  \eqref{eq:localcontribution} we use the shorthand notation $v_{ii}(u)= v_{ii}(x_i \cup u, x_i \cup
u)- v_{ii}(x_i \setminus u, x_i \setminus u)$, $v_{ij}^1(u)=
v_{ij}(x_i \cup u,x_j)-v_{ij}(x_i \setminus u,x_j)$, and $v_{ji}^2(u)=
v_{ji}(x_j,x_i \cup u)-v_{ji}(x_j,x_i \setminus u)$, $i \neq
j$. The Papangelou conditional intensity can then be written
as 
\[ \ld\{(u,i),y\} = \exp\biggl [\gm_{i0}^\top z_i(u)  + \gm_{ii}^\top
  v_{ii}(u)+\sum_{\substack{l=1\\l \neq i}}^p\Bigl\{   \gm_{il}^\top
  v_{il}^1(u) + \gm_{li}^\top v_{li}^2(u) \Bigr \}\biggr ].\]
In practice, we often have $z_i(u)=z_j(u)$ for all $i,j$. It is then clear
that adding the same constant to all $\gm_{i0}$ will not change the
log likelihood \eqref{eq:loglik}. This identifiability issue is
typically resolved by considering contrasts
$\beta_{i0}=\gm_{i0}-\gm_{p0}$ or, equivalently, fixing $\gm_{p0}=0$.

Suppose further that we choose the same type of interaction (e.g.\ Strauss hard-core with same
interaction ranges and hard-core distances) between $x_i$ and $x_l$ for
$i=1,\ldots,p$. Then  $v^1_{il}(u)=v^1_{jl}(u)$ for all $i,j$, and we
cannot identify the parameters $\gm_{il}$, $i=1,\ldots,p$ (and with similar considerations for the $\gm_{li}$). Again this
could be resolved by fixing one parameter, e.g.\
$\gm_{pl}=0$. Since we in practice  estimate the interaction ranges
from the given data, it is unlikely to encounter the situation
$v_{il}(u)=v_{jl}(u)$ for all $i \neq j$ which means that the
identifiability issue is less of a concern for the interaction
parameters $\gm_{il}$.

In case of symmetry, $v_{ij}(x_i,x_j)=v_{ji}(x_j,x_i)$, the parameters $\gm_{ij}$ and $\gm_{ji}$ are obviously not identifiable.
In general symmetry is not satisfied in which case one could consider
$\gm_{ij}\neq \gm_{ji}$.  One example is a multi-type Strauss process
with $R_{ij}<R_{ji}$. This becomes a kind of multiscale model where
$\gm_{ij}+\gm_{ji}$ is a parameter related to pairs of type $i$ and $j$ points
with interpoint distance less than $R_{ij}$ and $\gm_{ji}$ is a
parameter related to pairs with distance in $]R_{ij},R_{ji}]$.
Another example with $v_{ij}(x_i,x_j) \neq v_{ji}(x_j,x_i)$ is a multi-type Geyer saturation process  with
$c_{ij}=c_{ji}=1$ and $R_{ij}=R_{ji}$ (consider the case $x_i=\{u\}$
and $x_j=\{v,w\}$ where $u,v,w$ are all within distance
$R_{ij}=R_{ji}$ of each other). In the interest of model parameter parsimony one
may nevertheless prefer to restrict to the symmetric parametrization
$\gm_{ij}=\gm_{ji}$.

\subsection{Unbiasedness and approximate covariance matrix}\label{sec:unbiasednessandcovariance}

The basic requirement for consistency of the conditional pseudo
  likelihood estimator obtained by maximizing~\eqref{eq:loglik} is unbiasedness of the score function. Let $\lambda^\prime$ and $\Ld^\prime$ denote the derivative (i.e.\ gradient vector) of $\ld$ and $\Ld$ with respect to $\gm$. The conditional pseudo score of~\eqref{eq:loglik} is
\begin{equation}\label{eq:e}
e(\gm) = \frac{\dd \ell(\gm)}{\dd \gm}=\sum_{(u,i)\in Y_D} h\{ (u,i), Y \setminus (u,i) \}
\end{equation}
with
\[
h\{(u,i) , y\}=
 \frac{\ld^\prime \{(u,i), y \}}{\ld \{(u,i), y \}}
-
\frac{\Ld^\prime (u, y )}{\Ld (u, y )} = v\{(u,i),y \}-
E\{u,y \setminus (u,i)\}
\]
and
$E(u,y) = \sum_{l=1}^p v\{(u,l),y\}p\{(u,l),y\}$
for $(u,i)\in \R^d \times \mathcal M$ and $y\in \mathcal C$ with
$(u,i) \notin y$. Hence $E(u, y)$ can be viewed as a
conditional expectation of $v\{(u,I),y \}$ for a random index $I$
with $\PP(I=i)=p\{(u,i),y\}$, and $h\{(u,i),y \}$  can be viewed as 
a score residual vector. This suggests that $e(\gm)$ has mean zero
which can be formally verified using the GNZ formula~\eqref{eq:GNZ},
\begin{align*}
\EE \; e(\gm)   & = \EE \sum_{(u,i)\in Y_D}
h\{(u,i) , Y \setminus (u,i)\} = \EE \; \sum_{i=1}^p \; \int_D \; h\{(u,i),Y \}
  \ld\{(u,i),Y \} \; \dd u\\
&= \EE \int_D \left[\sum_{i=1}^p \ld^\prime\{(u,i),Y \} - \frac{\Ld^\prime(u,Y )}{\Ld(u,Y )} \sum_{i=1}^p \ld\{(u,i),Y \} \right] \dd u 
\\ &= \EE \int_D \Big\{ \Ld^\prime(u,Y )-\Ld^\prime(u,Y )\Big\} \; \dd u =0.
\end{align*}

Following the asymptotic results in Section~\ref{sec:asymp}, an
approximate covariance matrix of the estimator $\hat \gm$ is given in
terms of the sensitivity matrix $S(\gm)= - \EE \dd e(\gm)/\dd \gm^\top$ and the variance-covariance matrix  $\Sigma(\gm)= \Var\{e(\gm)\}$ of the score function.
We verify in Section~A in the supplementary material that the sensitivity matrix is
\begin{equation}\label{eq:S}
  S(\gamma) = \EE\int_D \Lambda(u,Y) V(u,Y) \dd u
  \end{equation}
where for $y \cap \{(u,1),\ldots,(u,p)\} =\emptyset$, $V(u,y) = E^2(u,y) -
  E(u,y)E(u,y)^\top$ can be viewed as a conditional variance with $E^2(u,y) =  \sum_{i=1}^{p}v\{(u,i),y\}v\{(u,i),y\}^{\top}p\{(u,i), y\}.$
We also obtain in Section~A in the supplementary material, under the finite range
assumption \eqref{eq:FR}, that the score covariance matrix is
\begin{align}
&\Sigma(\gamma)
=S(\gamma)+ \nonumber \\ &\EE\sum_{i,j=1}^p \! \int_D \! \int_{ D\cap B(u,R)}
\!\!\!\!\!\!\!\!\!\!\!\!\!\!\!\! \!\!\!
h\{(u,i), Y \cup (v,j)\} h\{(v,j), Y \cup (u,i)\}^{\top}  \ld[\{(u,i),(v,j)\},Y]
\dd u \dd v \label{eq:Sigma}
\end{align}
where we remind that
$\ld [\{(u,i),(v,j)\},Y] = \ld \{(u,i),Y\} \ld \{ (v,j), Y\cup (u,i) \}$.

The expressions \eqref{eq:S}-\eqref{eq:Sigma} cannot be evaluated due to the dependence on the unknown non-parametric factor $\phi_0$.
Following \citet{coeurjolly:rubak:13}, one may, however, apply GNZ~\eqref{eq:GNZ} and iterated GNZ~\eqref{eq:iteratedGNZ} backward to obtain the following unbiased estimators
\begin{align}
\hat S(\gm) &= \sum_{(u,i) \in Y_D}  V\{u,Y \setminus (u,i)\}  \label{eq:Sestgm}\\
\hat \Sigma(\gm)& = \sum_{\substack{(u,i),(v,j)\in Y_D:\\ \|u-v\| \le R}}^{\ne} h\{(u,i), Y\setminus (u,i)\}h\{(v,j), Y\setminus (v,j)\}^{\top}. \label{eq:Sigmaestgm} 
\end{align}

\subsection{Reparametrization}\label{sec:reparm}

As discussed in Section~\ref{sec:ident}, we may need to represent
the full parameter vector $\gm \in \R^k$ in terms of a lower dimensional parameter
vector $\beta \in \R^{k'}$, $k'<k$, i.e.\  $\gm=T(\beta)$ where $T: \R^{k'}\rightarrow
\R^{k}$ is a full rank linear mapping. Then an estimator $\hat \beta$
is obtained by maximizing the log likelihood
$\ell\{T(\beta)\}$. With an abuse of notation the resulting score function $e(\beta)=T^\top e \{T(\beta)\}$ is
still unbiased where we
also let $T$ denote the $k \times k'$ Jacobian matrix of the mapping
$T$. Continuing this abuse of notation, the resulting sensitivity matrix and score variance matrix become
\begin{equation}\label{eq:reparmsensvar} S(\beta)=T^\top S\{T(\beta)\}
  T \quad \text{and} \quad  \Sigma(\beta) = T^\top \Sigma\{ T(\beta) \}T, \end{equation}
and the approximate covariance matrix for $\hat \beta$ becomes $S(\beta)^{-1} \Sigma(\beta) S(\beta)^{-1}$. For the asymptotic results in Section~\ref{sec:asymp} we consider the reparametrized model with $\beta$ varying in an open convex subset of $\R^{k'}$.

\subsection{Kernel estimation of $\phi_0$}
\label{sec:Kernel}
Let $k$ be a probability density on $\R^d$ with support $B(0,\omega)$
for $\omega >0$. By GNZ,
\[ \EE \sum_{(u,i) \in Y} \frac{k(u-v)}{\exp \left[ \gm^\top
      v\{(u,i),y \setminus (u,i) \} \right]} = p\EE \int_{\R^d}
  \phi_0(u)k(u-v) \dd u \approx p\phi_0(v),\]
assuming $\phi_0(\cdot)$ is roughly constant on $B(v,\omega)$.
Hence we have the kernel estimator
\[ \hat \phi_0(v) = \frac{1}{p}\sum_{(u,i) \in Y} \frac{k(u-v)}{\exp \left[ \gm^\top v\{(u,i),y \setminus (u,i)\}\right]}.\]
It is an open question how to choose an optimal
  $\omega$ and the theoretical properties of the kernel estimator are
  unknown. We use the kernel estimator for purely exploratory purposes
  in Section~\ref{sec:banks}.

\section{Asymptotic results} \label{sec:asymp}

We consider a multi-type Markov model as defined in
Section~\ref{sec:semiparametricmodel}. We assume it
satisfies~\eqref{eq:LS} and \eqref{eq:FR} and thus there exists at least
one multi-type Markov point process on $\R^d\times \mathcal M$ which
encourages us to consider an increasing domain framework. Following Section~\ref{sec:reparm} we consider the reparametrization $\gm=T(\beta)$ and consider the properties of the estimator $\hat \beta \in \R^{k'}$.
We consider a sequence of increasing observation windows
  $(W_n)_{n\ge 1}$ and define as previously $D_n=W_n \ominus R$. We
  also add the index $n$ to emphasize the dependence of $\hat \beta_n$, $\ell_n$, $e_n$, $S_n$, and $\Sigma_n$ on the observation window $W_n$.

For our asymptotic results we need the following quite mild conditions (see discussion of conditions in the final paragraph of this section).
\begin{enumerate}
\renewcommand\labelenumi{C.\arabic{enumi}}
\renewcommand\theenumi\labelenumi

\item  The sequence of eroded observation windows $D_n$ is increasing with
  $\lim_{n \rightarrow \infty}|D_n|=~\infty$.  \label{C:Dn}
\item The Papangelou conditional intensity function has
  the log-linear specification given by \eqref{eq:PapangelouMarked2}-\eqref{eq:localcontribution}. A reparametrization
  $\gamma=T(\beta)$ is used with $\beta \in B \subseteq \R^{k^\prime}$ where $T$
  is a full rank linear mapping and $B$ is an open convex subset
  of $\R^{k^\prime}$. The
  property~\eqref{eq:LS} is satisfied for the true value of $\gamma
  =T(\beta)$. The finite range property \eqref{eq:FR} is satisfied for the conditional intensity (and hence for $v\{(u,i),y\}$) for some $0< R < \infty$.  \label{C:model}
\item \label{C:boundedinteraction} Recall the notation in
    Section~\ref{sec:ident} for the interaction components
    of $v\{(u,i),y\}$. We assume that there exist $c_1>0$
and $c_2\ge 0$ such that for any $(u,i) \in
  \R^d \times \M$ and $y$ with $\ld\{(u,i),y\}>0$,  $\sup_{u \in \R^d}\|z_i(u)\| < c_1$, $i=1,\ldots,p$, and
\begin{align}\label{eq:vell}
  &\|v_{ii}(u)\|   \le c_1 N_i\{ B(u,R)\}^{c_2}, \quad \|v^1_{ij}(u)\| \le c_1 N_j\{ B(u,R)\}^{c_2} \nonumber\\
  &\|v^2_{ji}(u)\| \le c_1 N_j\{ B(u,R)\}^{c_2}.
\end{align}
\item The scaled limiting sensitivity  and score variance matrices satisfy
$$ \varphi_S^*=\displaystyle \liminf_{n\to \infty} \;
    \varphi_{\min}\big\{|D_n|^{-1}  S_n(\beta) \big\}>0
    \text{ and }
    \varphi_\Sigma^*=\displaystyle \liminf_{n\to \infty} \;
    \varphi_{\min}\big\{|D_n|^{-1}  \Sigma_n(\beta) \big\}>0,
    $$
where $\varphi_{\min}(M)$ denotes the smallest eigenvalue of a matrix $M$.
\label{C:mat}
\end{enumerate}
The following result states the asymptotic behavior of the
conditional pseudo likelihood estimator.

\begin{theorem} \label{thm:asymp}
Under the assumptions \ref{C:Dn}-\ref{C:mat}, the following holds as $n\to \infty$.\\
$(i)$ There exists a sequence of local maxima $\hat{\beta}_n$ of $\ell_n(\beta)$ such that $\|\hat \beta_n - \beta\| = O_{\PP}(|D_n|^{-1/2})$.\\
$(ii)$ The sequence $\{\hat{\beta}_n\}_{n \ge 1}$ is asymptotically normal,
\begin{equation} \label{eq:normality}
   \Sigma_n(\beta)^{-1/2}    S_n(\beta) (\hat \beta_n-\beta) \stackrel{d}{\rightarrow} N(0,I_{k^\prime}),
\end{equation}
where $ \Sigma_n(\beta)$ and $ S_n(\beta)$ are given by~\eqref{eq:reparmsensvar}.
\end{theorem}

The proofs of these results are given in Section~B in the supplementary material. By~\eqref{eq:normality}, the covariance
matrix of $\hat \beta_n$ can be approximated by $ V_n(\beta) =
 S_n^{-1}(\beta)  \Sigma_n(\beta) 
S_n(\beta)^{-1}$. In practice we replace $S_n$ and $V_n$ by the estimates \eqref{eq:Sestgm} and
\eqref{eq:Sigmaestgm} and  plug in $\hat \beta_n$ for $\beta$. This
allows us to derive asymptotic confidence intervals for the components of $\beta$.

Consider the assumptions \ref{C:model}-\ref{C:boundedinteraction} for the models mentioned
in Section~\ref{sec:semiparametricmodel} for which
properties~\eqref{eq:LS}-\eqref{eq:FR} are valid as already
discussed when covariates and the non-parametric factor are bounded. 
The condition \eqref{eq:vell} is also satisfied for the aforementioned models with $c_2=0$ for the multi-type Strauss hard-core  interaction (with $r_{ij}>0$), the multi-type Geyer saturation interaction, and  the multi-type area-interaction, and with $c_2=1$ for the multi-type Strauss interaction. Assumption \ref{C:mat}
depends on the behaviour of the non-parametric factor and the
covariates on the whole of $\R^d$ and thus cannot be verified.
However, for \ref{C:mat} to hold, it is required that the
  intersection of $D_n$ with the support of $\phi_0$ grows at the same
  rate as $D_n$.

\section{Simulation Study}
We demonstrate the usefulness of the conditional pseudo
  likelihood approach by a simulation study considering a semi-parametric multi-type Markov model with $\mathcal
M=\{1,2,3\}$, and first order
effects depending on an intercept and one covariate
$z=\{z(u)\}_{u \in W}$ common to all three types of  points.
The baseline function $\varphi_0$ and the
covariate $z$ are obtained as realizations of two independent Gaussian
random fields (GRF). Both are simulated with an exponential covariance
function, $C(u,v) = \sigma^{2}\exp( -\norm{u-v}/\phi)$. For $\varphi_0$ we use  a GRF (truncated
at zero if needed) with mean
350, $\sigma^{2} = 900$, and $\phi = 0.1$, and for $z$ we use a zero mean GRF with
$\sigma^{2}=0.2$ and $\phi = 0.1$.

For the interaction part, we consider an
inhomogeneous multi-type Poisson process, a multi-Strauss process, and a
multi-Geyer
saturation process (for brevity, Poisson, Strauss, and Geyer in the following). For all three processes, 1800 replications are
simulated on $W_1=[0,1]^2$ and $W_2=[0,2]^2$ using the R package
\texttt{spatstat} \citep{baddeley:rubak:turner:15}. From each
simulation we obtain parameter estimates as well as  estimated standard errors
and approximate 95\% confidence intervals based on the asymptotic
results in Section~\ref{sec:asymp}. The number 1800 yields a
Monte-Carlo standard error of 0.005 for the estimated coverage rates provided the true coverage rate is 95\%.
Single realizations of $\varphi_0$ and $z$ are simulated on
$W_2$ and used across all replications. For all three
processes we use parameters $\gm_{10,2} = 1/2$,
$\gm_{20,2} = -1/2$, $\gm_{30,2} = 0$ for the covariate $z$. For the intercepts we use
$\gm_{10,1} = \gm_{20,1} = \gm_{30,1} = 0$ for Poisson, $\gm_{10,1} = \gm_{20,1} = \gm_{30,1} = \log 1.6$ for Strauss, and $\gm_{10,1} = \log (1.3/1.4)$, $\gm_{20,1} = \log (1.3/1.6)$, $\gm_{30,1}=\log 1.3$ for Geyer.

For all three
processes, we use a within-type interaction range of 0.02 and a
between-type interaction range of 0.04 for both the simulation (where
relevant) and model fitting. For Strauss, we use
interaction parameters $\gm_{ii} = \log 0.8$ and $\gm_{ij} = \log
0.9$, $i \neq j$. No function for simulating Geyer multi-type
processes with between-type interaction exists in \texttt{spatstat}.
Instead we simulate three independent
single-type saturation processes, and thus no between-type interaction
is present. We choose within-type interaction parameters $\gm_{11} = \log 1.1$, $\gm_{22} =
\log 1.2$, $\gm_{33} = \log 0.8$, and a common saturation parameter 10 for
all three types of points. The expected numbers of each type of points
across the three models fall in the range 360-385 in case of $W_1$ and
in the range 1417-1519 in case of $W_2$.

\begin{figure}[t]
\centering
\caption{ {\fontsize{10pt}{12pt}\selectfont Kernel density plots of estimates of the parameters
  $\bt_{10,1}, \beta_{10,2}, \gm_{11},$ and $\gm_{12}$
  for  simulations of the multi-type Strauss process on $W_1=[0,1]^2$. The  true parameter value and the mean parameter estimate are shown with
  blue and red vertical bars.} }\label{fig:kdestrauss}
  \includegraphics[width=0.8\textwidth]{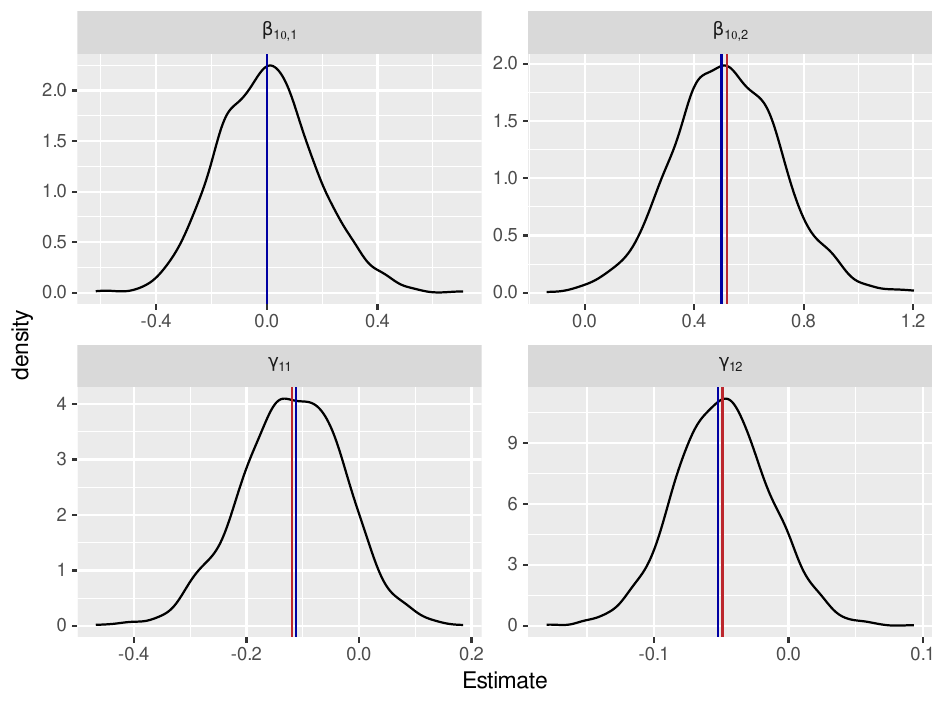}
\end{figure}

The Poisson process is nested within the Strauss
process (with the interaction parameters equal to zero) and we fit the
Strauss process for both the
Poisson and the Strauss simulations.  We
impose symmetry on the interaction parameters $\gm_{ij}=\gm_{ji}$
for any $i\neq j$ and we take the third type as the reference type
for the intercept and covariate parameters, i.e.\ we estimate
$\beta_{i0}=\gm_{i0}-\gm_{30} \in \RR^2$, $i=1,2$.
Figure~\ref{fig:kdestrauss} shows a few representative kernel
density plots of parameter estimates obtained from simulations of the
Strauss process in case of the window $W_1$.
The plots show that the estimates are close to being normally distributed with
negligible bias. Further plots (not shown) confirm this for the
remaining parameter estimates under Poisson, Strauss, and Geyer simulations.

\begin{figure}[t]
\centering
\caption{{\fontsize{10pt}{12pt}\selectfont Standard errors (points and solid lines) and means of
  estimated standard errors (triangles and dashed lines) for Poisson
  (red), Strauss (blue), and Geyer (teal) and windows $W_1$ (upper curves) and
  $W_2$ (lower curves). Curves are added for easier visual interpretation.}}\label{fig:stderrpoissonstrauss}
\centering
\includegraphics[width=0.8\textwidth]{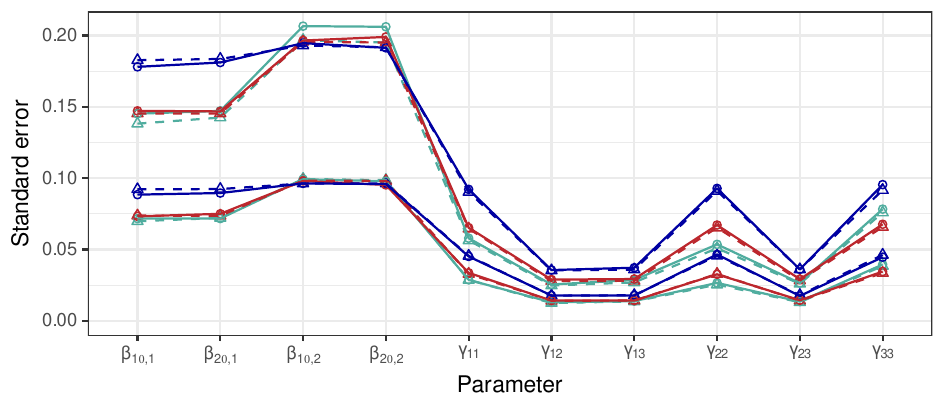}
\end{figure}

Figure~\ref{fig:stderrpoissonstrauss} shows standard errors of the
parameter estimates as well as means of estimated standard errors for all three models.
For Poisson and Strauss, the means of
estimated standard errors are in good agreement with the actual
standard errors. For Geyer, the standard errors for the parameters associated with the spatial
covariates and intercepts are somewhat underestimated on $W_1$, while all standard errors are
well-estimated on $W_2$. In accordance with the asymptotic results, the ratios
between standard errors for $W_2$ and $W_1$ are close to $\sqrt{|W_2|/|W_1|}=2$.

\begin{figure}[t]
  \centering
\caption{{\fontsize{10pt}{12pt}\selectfont Coverage rates for nominal 95\% confidence intervals in case of $W_1$ (upper plot) and $W_2$ (lower plot). Red,
   blue, and teal for simulations of Poisson, Strauss, and
  Geyer processes. 95\% Monte
  Carlo probability intervals for the coverage rates are
  shown with black dashed lines. Curves are added for easier visual interpretation.}}\label{fig:coverages}
\centering
\includegraphics[width=0.8\textwidth]{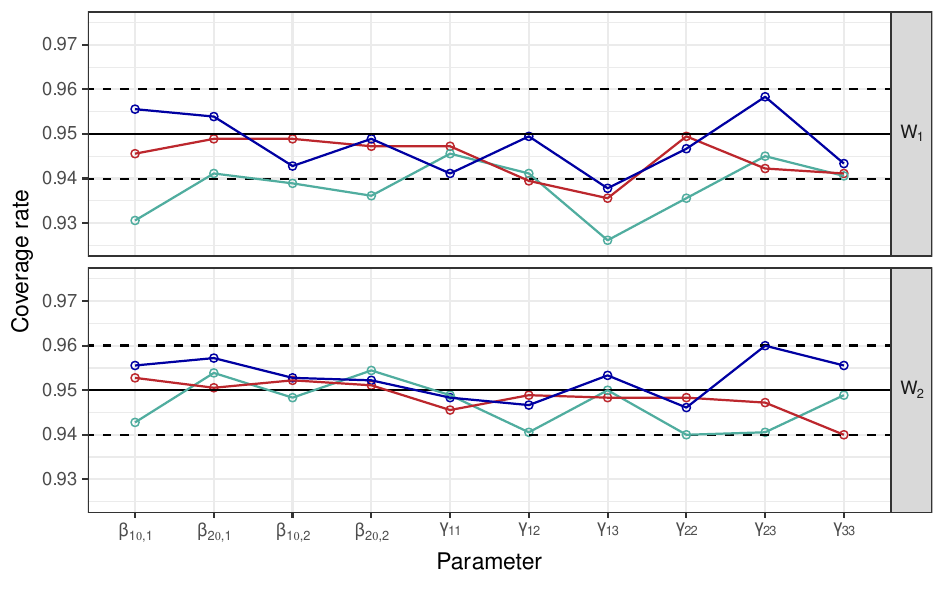}
 \end{figure}

Figure~\ref{fig:coverages} shows estimates of coverage rates for approximate 95\%
confidence intervals based on asymptotic normality as well as 95\%
Monte Carlo probability intervals for the
coverage rates assuming that the true coverage rate is 95\%. On $W_1$, coverage rates are very close to
the nominal 95\% for Poisson and Strauss and a bit too low for Geyer. On $W_2$, coverage rates are close to the nominal level for all models.

\section{Clustering of French banks}\label{sec:banks}

Most French banks can be divided into two types: cooperative and
lucrative. Cooperative banks originated within local communities with
the purpose of providing loans to local retail customers and
moderately sized enterprises. Lucrative banks are limited liability
companies typically founded within larger cities. We consider a
dataset of locations of 11,244 cooperative banks and 3,448 lucrative
banks recently collected by \cite{artis2025marked}, see Figure \ref{fig:Banks}. The bank locations are aggregated from three different cooperative banks and two different lucrative banks. As of 01/01/2023, these banks cover 92\% of French clients.

In addition to the bank locations,
socio-economic variables and estimated population density are
available for 1681 administrative units covering France
\citep{artis2025marked}. Detailed definitions of these covariates are
given in Section~C in the supplementary
material (see also Figure~\ref{fig:BankCI}). We treat these variables as piece-wise constant covariates with constant value
within administrative units.

We model the data using a bivariate Geyer
saturation model which allows for both positive and negative
interactions within and between the two types of banks. In the context of banks, the Geyer saturation parameter  models an effect
of diminishing returns after reaching a ceiling - e.g.\ a possible positive
effect on the conditional intensity from adding a neighbouring bank
vanishes when a neighbourhood is already saturated with banks.
Our first aim is to study possible differences
regarding the association between the bank locations and the
socio-economic variables. While the structures of the spatial patterns of the two types
of banks may differ, it is clear that the intensities of the banks are
strongly dependent on urbanization. If this trend is of a similar form for cooperative and lucrative banks, it can be accounted for by the non-parametric
factor in our model. Urbanization is likely closely related to
population density so that including both the non-parametric factor as
well as population density might seem superfluous. However, when
population density is included as a covariate it only explains
discrepancies between the conditional intensities of one type of bank
relative to the other. Thus the non-parametric factor may capture absolute effects of population density among other spatial factors common to both types of banks.
Secondly, we assess possible unexplained interactions  within and between the
types of banks by considering estimates of the interaction
parameters. 

As mentioned in Section~\ref{sec:ident}, it is not possible to
identify the absolute effects of the spatial covariates on the
intensities of each type of banks.
We use the lucrative banks as a reference and estimate the
effects of the spatial covariates on the cooperative banks relative to
the lucrative banks. It is on the other hand possible to identify all interaction
effects provided not all interaction ranges are identical. We use a common value $R_w=R_{11}=R_{22}$ for the within-type interaction ranges and another value $R_b=R_{12}=R_{21}$ for the between-type interaction ranges. We moreover use a common saturation parameter $c$ for all combinations of types of banks. We
estimate $R_w$, $R_b$, and $c$ by fitting the
model over a grid of combinations of interaction ranges and saturation
parameter and choosing the combination  with the largest conditional
pseudo likelihood. This results in values $R_w = 0.006$, $R_b = 0.004$,
and $c=4$. The bank locations are given in terms
of longitude and latitude, and the interaction ranges correspond to
approximately 300 and 200 metres respectively. To avoid edge effects,
we erode the observation window by a distance of twice the largest interaction range included in the grid search, in this case $0.02$.

Figure~\ref{fig:BankCI} illustrates the parameter estimates. For
\texttt{activity} and \texttt{propindustry}, the confidence intervals
include zero. The confidence intervals for \texttt{prop.age},
\texttt{pop.growth}, and \texttt{proppublic} are confined to positive
parameter values, and the confidence intervals for \texttt{income.decile},
\texttt{prop.firms}, \texttt{proptrade}, and \texttt{log-density} to
negative parameter values. This indicates that administrative units
with a higher proportion of younger people relative to older people, greater recent population
growth, and a high proportion of public-sector jobs tend to have a higher proportion of cooperative banks. Conversely, administrative units with higher income inequality, a greater proportion of small businesses, a higher proportion of jobs in the trade sector, and higher population density tend to have a higher proportion of lucrative banks. A plot showing the fitted probability that a bank is cooperative conditional on the rest of the point pattern can be found in Figure~S1 in the supplementary material.

Both estimated within-type interactions are negative, while the estimated
between-type interaction is positive.
It is well-known in the
economics literature that banks tend to cluster
\citep{Lord:Wright:81, Jayaratne:97}. One might hypothesize many
potential explanations for this phenomenon, such as proximity to a
qualified labour-pool or beneficial infrastructure. The findings of
\cite{Lord:Wright:81} indicate that a contributing factor might be so-called
\textit{rational herding}. This is a phenomenon that occurs when individuals
or firms believe that others possess more information than themselves. As a
result, firms may mimic one another, even if the resulting outcome is not optimal.
On the other hand, banks also compete with each other,
which can explain negative interactions.
Between these effects, the semi-parametric Markov model
suggests that after taking spatial covariates and the common trends embedded in the non-parametric factor into account, competition dominates for banks of the same type. The positive cross-type interaction between lucrative and cooperative banks might indicate that they avoid competition by targeting different customers, while retaining some of the effects that lead to clustering.
\begin{figure}[t]
\centering
\caption{{\fontsize{10pt}{12pt}\selectfont Parameter estimates (points) and confidence intervals (lines) from the semi-parametric Markov model. Parameters associated with the spatial covariates represent the effect on the cooperative banks relative to the effect of the lucrative banks. The 95\% confidence intervals (CI) are calculated using the standard errors estimated based on the asymptotic results.}}
\label{fig:BankCI}
\includegraphics[width=0.8\textwidth]{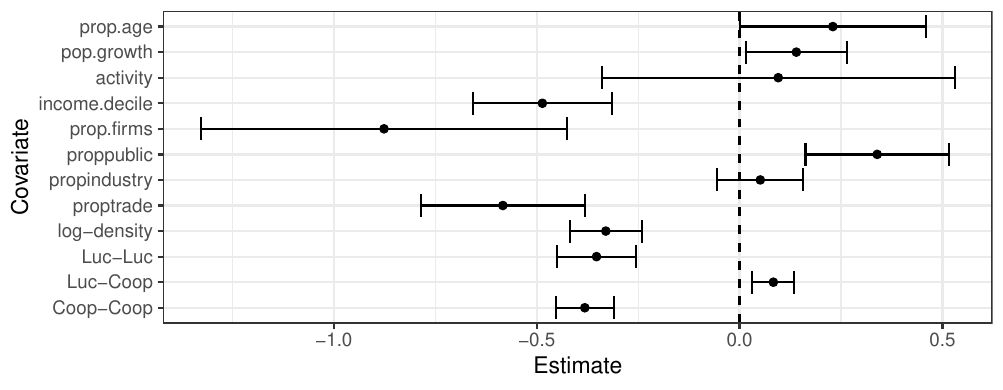}
\vspace*{-1cm}
\end{figure}

\begin{figure}[!ht]
  \caption{{\fontsize{10pt}{12pt}\selectfont Estimate of $\log \phi_0$ obtained with band width $\omega =
    0.25$ and an Epanechnikov kernel averaged over administrative
      units (left) along with log-population density (right).}}
  \centering
 \includegraphics[width=0.8\textwidth]{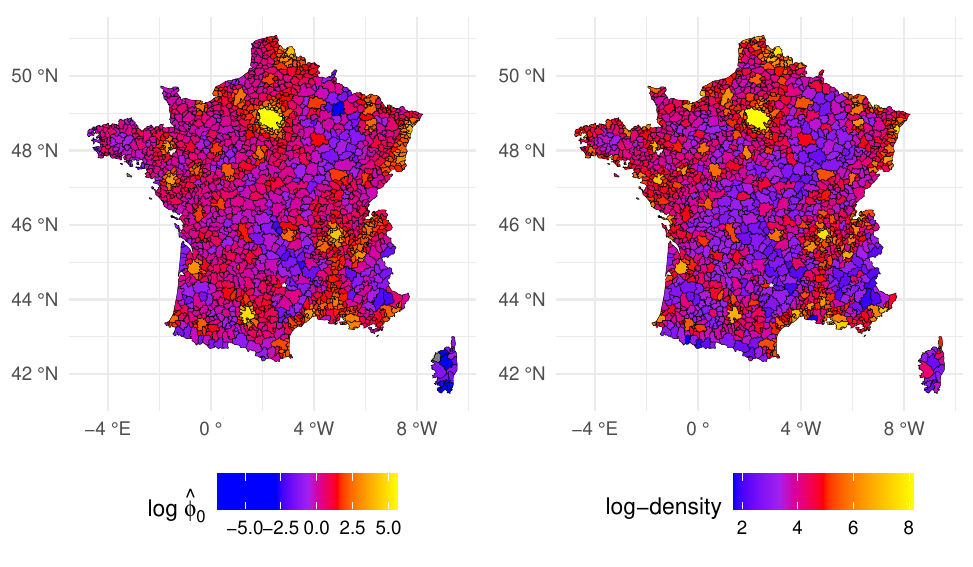}
 \label{fig:Kernel_est_phi0}
\end{figure}

Finally, we estimate $\phi_0$ as described in Section
\ref{sec:Kernel}. Figure~\ref{fig:Kernel_est_phi0} shows a plot of
$\log \hat{\phi}_0$ across France with band width $\omega = 0.25$
chosen large enough to avoid zero estimates of $\phi_0$. The
  estimates of $\phi_0(u)$ for locations $u$ in France were
  averaged within administrative regions to ease visual comparison
  with the log population density, also shown in Figure
  \ref{fig:Kernel_est_phi0}. Visual inspection and
  a correlation of 0.73 between  $\log \hat{\varphi}_0$ and log
  population density suggest that the non-parametric factor $\phi_0$
  is effective in picking up effects of population density common to
  both types of banks. This is reassuring for further applications where
  important spatial variables might be unobserved. Moreover, although
  $\phi_0$ seems to be strongly associated with population density, it may capture effects of further unobserved spatial variables as well in the current bank example.

\section{Discussion}

In this paper we significantly expand the toolbox of spatial
statistics by introducing a semi-parametric approach for inhomogeneous multi-type
Markov point processes. We do this by introducing a non-parametric
factor adjusting for  unobserved spatial variables that might
otherwise confound parameter estimation. We moreover introduce a
computationally efficient conditional pseudo likelihood method for
parameter estimation. Asymptotic results and simulation studies demonstrate the good properties of the resulting parameter estimates. The application to French banks shows how the model can be used to decompose spatial patterns according to impacts of spatial covariates and within- and between-type interactions where the latter can be linked to economic theories of industry co-agglomeration.

Model assessment is an open problem for the semi-parametric Markov
model. Commonly used point process residuals
\citep{baddeley2005residual} and second order point process
summary statistics like the $K$-function require a consistent estimator of the
intensity function. Plugging in the kernel estimate of the
non-parametric factor is not sufficient since this kernel estimate is not consistent. In
Section~\ref{sec:unbiasednessandcovariance} we discussed zero-mean
conditional pseudo score residuals whose properties  might be explored
further following ideas in \cite{coeurjolly2013residuals}.

%
%
%

\section{Acknowledgments}
Ib T.\ Jensen and R.\ Waagepetersen were supported by grants VIL57389, Villum Fonden, and
NNF23OC0084252, Novo Nordisk Foundation.

\bibliographystyle{abbrvnat}
\bibliography{Bib/mybib}

\end{document}